\documentclass[%        	Class options:
aps,%                   	American Physical Society
prd,%                   	Physical Review D
superscriptaddress,%    	Authors' addresses linked with superscripts
nofootinbib,%           	Does not treat footnotes as references
floatfix]%              	Fixes float errors
{revtex4}%              	REVTEX 4 Package used
\usepackage{graphicx,%  	Default Latex 2eps package for embedding figures
longtable,%             	Useful for long table
color,%					Colours
bm}%						Bold math symbols
\begin{document}
%======================================================================================
\title{Addendum to: Global constraints on absolute neutrino masses and their ordering}
%--------------------------------------------------------------------------------------
%
\author{        	Francesco~Capozzi}
\affiliation{   	Max-Planck-Institut f\"ur Physik (Werner-Heisenberg-Institut), F\"ohringer Ring 6, 80805 M\"unchen, Germany}
\author{			Eleonora Di Valentino}
\affiliation{	Jodrell Bank Center for Astrophysics, 
School of Physics and Astronomy,
University of Manchester, Oxford Road, Manchester, M13 9PL, UK}
\author{        	Eligio~Lisi}
\affiliation{   	Istituto Nazionale di Fisica Nucleare, Sezione di Bari, %\\
               	Via Orabona 4, 70126 Bari, Italy}
\author{        	Antonio~Marrone}
\affiliation{   	Dipartimento Interateneo di Fisica ``Michelangelo Merlin,'' %\\
               	Via Amendola 173, 70126 Bari, Italy}%
\affiliation{   	Istituto Nazionale di Fisica Nucleare, Sezione di Bari, %\\
               	Via Orabona 4, 70126 Bari, Italy}
\author{			Alessandro~Melchiorri}
\affiliation{	Dipartimento di Fisica, Universit{\`a} di Roma ``La Sapienza,'' P.le Aldo Moro 2, 00185 Rome, Italy}
\affiliation{   	Istituto Nazionale di Fisica Nucleare, Sezione di Roma~I, %\\
               	P.le Aldo Moro 2, 00185 Rome, Italy}
\author{        	Antonio~Palazzo}
\affiliation{   	Dipartimento Interateneo di Fisica ``Michelangelo Merlin,'' %\\
               	Via Amendola 173, 70126 Bari, Italy}%
\affiliation{   	Istituto Nazionale di Fisica Nucleare, Sezione di Bari, %\\
               	Via Orabona 4, 70126 Bari, Italy}
\begin{abstract}%.......................................................................
\medskip
We revisit our previous work [Phys.\ Rev. D {\bf 95}, 096014 (2017)] where neutrino oscillation and nonoscillation data were analyzed in the standard framework with three neutrino families, in order to constrain their absolute masses and to probe their ordering (either normal, NO, or inverted, IO). We include updated oscillation results to discuss best fits and allowed ranges for the two squared mass differences $\delta m^2$ and $\Delta m^2$, the three mixing angles $\theta_{12}$, $\theta_{23}$ and $\theta_{13}$, as well as constraints on the CP-violating phase $\delta$, plus significant indications in favor of NO {\em vs\/} IO at the level of $\Delta\chi^2=10.0$. We then consider nonoscillation data from beta decay, from neutrinoless double beta decay (if neutrinos are Majorana), and from various cosmological 
input variants (in the data or the model) leading to results dubbed as default, aggressive, and conservative. In the default option, we obtain from nonoscillation data an 
extra contribution $\Delta\chi^2\simeq 2.2$ in favor of NO, and an upper bound on the sum of neutrino masses $\Sigma < 0.15$~eV at $2\sigma$; both results ---dominated by cosmology--- can be strengthened or weakened by using more aggressive or conservative options, respectively. Taking into account such variations, we find that the combination of all (oscillation and nonoscillation) neutrino data favors NO at the level of $3.2$--$3.7\sigma$, and that $\Sigma$ is constrained at the $2\sigma$ level within $\Sigma< 0.12-0.69$~eV. The upper edge of this allowed range corresponds to an effective $\beta$-decay neutrino mass $m_\beta\simeq \Sigma/3\simeq 0.23$~eV, at the sensitivity frontier of
the KATRIN experiment. 
\medskip
\end{abstract}%.........................................................................
\maketitle

%%%%%%%%%%%%%%%%%%%%%%%%%%%%%%%%%%%%%%%%%%%%%%%
\section{Introduction}
\label{Sec:Intro}
%%%%%%%%%%%%%%%%%%%%%%%%%%%%%%%%%%%%%%%%%%%%%%%

In a previous work \cite{Capozzi:2017ipn} 
 we have discussed in detail the constraints on absolute neutrino masses and their ordering arising from a global analysis of world $\nu$ data available in 2017, within the standard framework for three neutrino families ($3\nu$). We think it useful to reassess those findings by using more recent experimental results. In particular, we provide updated estimates of mass-mixing oscillation parameters, discuss statistically significant indications in favor of the so-called ``normal mass ordering'' from (non)oscillation data, and present constraints on absolute $\nu$ masses, involving different combinations of cosmological data and models.  

This Addendum is structured as follows. In Sec.~\ref{Sec:Param} we briefly recall the basic $3\nu$ parameters and observables, and  the methodology adopted in our analysis. In Sec.~\ref{Sec:Osc} we present updated oscillation data and parameter constraints, including indications in favor of normal ordering. In Sec.~\ref{Sec:Nosc} we discuss recent nonoscillation results from single and double beta decay and from cosmology, with emphasis on the latter---in view of possible departures from ``default'' choices towards more ``aggressive'' or ``conservative'' options, altering the impact on the mass ordering and on absolute $\nu$ masses.  Taking into account these variants, in the final Sec.~\ref{Sec:Conc}
we find upper bounds on the sum of neutrino masses $\Sigma$ in the range $0.12$--$0.69$~eV at $2\sigma$, and 
an overall indication for normal ordering at the level of $3.2$--$3.7\sigma$. 

%%%%%%%%%%%%%%%%%%%%%%%%%%%%%%%%%%%%%%%%%%%%%%%
\section{Parameters, observables and methodology}
\label{Sec:Param}
%%%%%%%%%%%%%%%%%%%%%%%%%%%%%%%%%%%%%%%%%%%%%%%

We adopt the standard $3\nu$ framework \cite{Tanabashi:2018oca},
where the three flavor states $\nu_\alpha$ $(\alpha=e,\,\mu,\,\tau)$
are linear combinations of three massive states $\nu_i$ $(i=1,\,2,\,3)$. The main parameters are the three $\nu$ masses $m_i$, the three mixing angles $\theta_{ij}$ and the CP-violating phase $\delta$, supplemented by two extra phases in the case of Majorana neutrinos. Neutrino propagation in matter greatly enriches the phenomenology related to these parameters. See \cite{PDGreview} and references therein.

Concerning neutrino oscillations, their amplitudes and frequencies are sensitive to (at least one) of the angles $\theta_{ij}$ and of the  
squared mass differences $\Delta m^2_{ij}$, respectively.   
We define $\delta m^2 = m^2_2-m^2_1>0$ and 
$\Delta m^2 = m^2_3-(m^2_2+m^2_1)/2$, 
where $\Delta m^2>0$ or $<0$ in the so-called normal ordering (NO) or inverted ordering 
(IO) for the neutrino mass spectrum, respectively. The channel $\nu_\mu\to \nu_e$ provides some sensitivity to $\delta$, as well as to $\pm\Delta m^2$ via matter effects. 
In the analysis, we start with the minimal data set sensitive to all the oscillation parameters $(\delta m^2,\,\pm\Delta m^2,\,\theta_{ij},\,\delta)$, as provided by the combination of solar, KamLAND and long-baseline (LBL) accelerator data. By adding short-baseline (SBL) reactor data, one constrains directly the pair $(\pm\Delta m^2,\,\theta_{13})$ and, to some extent, the parameters $(\theta_{23},\,\delta)$ via covariances in the fit. Finally, by adding atmospheric data, one further increases the sensitivity to $(\pm \Delta m^2,\theta_{23},\,\delta)$. Oscillation data do not constrain absolute $\nu$ masses, but 
reduce the phase space of nonoscillation observables. 

Nonoscillation observables include: the sum of $\nu$ masses $\Sigma$ probed by cosmology, the effective mass $m_\beta$ probed in beta decay, and the effective mass $m_{\beta\beta}$ probed in neutrinoless double beta decay (if neutrinos are Majorana); see 
\cite{PDGreview,Capozzi:2017ipn}
 for definitions. Concerning $\Sigma$ we remark that, as advocated in \cite{Capozzi:2017ipn}, our analysis of cosmological data accounts for three different masses $m_i$ (as dictated by the nonzero values of $\delta m^2$ and $\pm \Delta m^2$) and does not assume the degenerate-mass approximation ($m_1=m_2=m_3=\Sigma/3$). Our approach allows to correctly estimate the NO--IO differences at relatively small values of $\Sigma$, and to recover the degenerate case in the limit of high $\Sigma$ (where NO and IO converge).
 
Best fits and constraints on the $\nu$ parameters are obtained via a $\chi^2$ 
approach. 
Single-parameter bounds are obtained by projecting away all the others, so that $N_\sigma = \sqrt{\Delta \chi^2}$ defines the distance from the best fit in standard deviation units.  
This metric can also be applied to test the discrete hypotheses of NO vs IO \cite{PDGreview,Uchida:2018wup}.
In the analysis of cosmological data, likelihoods are transformed into effective $\chi^2$ values as described in \cite{Capozzi:2017ipn}.

%%%%%%%%%%%%%%%%%%%%%%%%%%%%%%%%%%%%%%%%%%%%%%%
\section{Oscillation data and constraints}
\label{Sec:Osc}
%%%%%%%%%%%%%%%%%%%%%%%%%%%%%%%%%%%%%%%%%%%%%%%

Concerning oscillation data, the analysis presented in \cite{Capozzi:2017ipn} 
has been updated in a subsequent review \cite{Capozzi:2018ubv}. With respect to 
\cite{Capozzi:2018ubv}, we include LBL accelerator data as published by the Tokai-to-Kamioka (T2K) experiment  
\cite{Abe:2019vii} and by the NuMI Off-axis $\nu_e$ Appearance (NOvA) experiment  
\cite{Acero:2019ksn}. Concerning SBL reactor data, we include the most recent results from the Daya Bay experiment 
\cite{Adey:2018zwh} 
and the Reactor Experiment for Neutrino Oscillation (RENO) \cite{Bak:2018ydk}; they 
dominate the current constraints on $\theta_{13}$ and, at the same time, provide a measurement of $\Delta m^2$ independent from accelerator and atmospheric data. 
In the analysis of Gallium solar neutrino data (GALLEX-GNO and SAGE) we account for the reevaluation of the $\nu_e$-Ga cross-section in \cite{Kostensalo:2019vmv}, although its effect on the fit turns out to be tiny. 

For the sake of completeness, we also mention some recent results that are not included in this work but might be eventually considered in the future: $(i)$ SAGE data with additional exposure have been preliminary reported in 
\cite{Gavrin2019}, but have not been published yet (to our knowledge); $(ii)$ new Double Chooz measurements of $\theta_{13}$ have been released in \cite{DoubleChooz:2019qbj}, but assuming a prior on $\Delta m^2$ that prevents inclusion in a global fit; $(iii)$ additional atmospheric $\nu$ results have been reported by the Super-Kamiokande (SK) \cite{Jiang:2019xwn} and IceCube Deep Core (IC-DC) \cite{Aartsen:2019eht} experiments, but they have not been cast (yet) in a format that can be reproduced or effectively used outside the collaborations --- hence we continue to use the previous $\chi^2$ maps from SK and IC-DC as described in \cite{Capozzi:2018ubv}.

The results of our global analysis of oscillation data  are reported in Table~\ref{Tab:Synopsis}, in terms of allowed ranges at 1, 2 and $3\sigma$ for each oscillation parameter (the other parameters being marginalized away), for the separate cases of NO and IO. The last column shows the formal $1\sigma$ accuracy reached for each parameter. It is interesting to notice that  the parameter $\theta_{23}$ is now being constrained with an overall fractional accuracy approaching that of $\theta_{12}$, although its best fit  remains somewhat unstable, due to the quasi-degeneracy of the $\theta_{23}$ octants \cite{Fogli:1996pv}. Also, if one takes the current constraints on $\delta$ at face value, then this parameter is already being ``measured'' with $O(10)\%$ accuracy, around a best-fit value suggestive of nearly maximal CP violation ($\delta \sim 3\pi/2$). However, the CP-conserving value $\delta =\pi $ is still allowed at $\sim\! 1.6 \sigma$ (i.e., at $\sim \!90\%$~C.L.) in our global fit, where the CP-violating hint coming from T2K data \cite{Abe:2019vii} is somewhat diluted in combination with current NOvA data \cite{Acero:2019ksn}.

%===========================================================================
\begin{table}[b]
\centering
\resizebox{0.85\textwidth}{!}{\begin{minipage}{\textwidth}
%\captionsetup{width=.9\textwidth}
\caption{\label{Tab:Synopsis} %\footnotesize 
Global $3\nu$ analysis of oscillation data, in terms of best-fit values and allowed ranges at $N_\sigma=1$, 2, 3 for the mass-mixing parameters, in either NO or  IO. The last column shows the formal ``$1\sigma$ accuracy'' for each parameter, defined as 1/6 of the $3\sigma$ range, divided by the best-fit value (in percent). 
We recall that 
$\Delta m^2=m^2_3-{(m^2_1+m^2_2})/2$ and $\delta/\pi\in [0,\,2]$ (cyclic).}
%\vspace*{0mm}
%\centering
%\resizebox{.9\textwidth}{!}{
\begin{ruledtabular}
\begin{tabular}{lcccccc}
%\hline\hline
Parameter & Ordering & Best fit & $1\sigma$ range & $2\sigma$ range & $3\sigma$ range & ``$1\sigma$'' (\%) \\
\hline%---------------------------------------------------------------------
$\delta m^2/10^{-5}~\mathrm{eV}^2 $ & NO & 7.34 & 7.20 -- 7.51 & 7.05 -- 7.69 & 6.92 -- 7.90 & 2.2 \\
							  		& IO & 7.34 & 7.20 -- 7.51 & 7.05 -- 7.69 & 6.92 -- 7.91 & 2.2 \\
\hline%---------------------------------------------------------------------
$\sin^2 \theta_{12}/10^{-1}$ & NO & 3.05 & 2.92 -- 3.19 & 2.78 -- 3.32 & 2.65 -- 3.47 & 4.5 \\
							 & IO & 3.03 & 2.90 -- 3.17 & 2.77 -- 3.31 & 2.64 -- 3.45 & 4.5 \\
\hline%---------------------------------------------------------------------
$|\Delta m^2|/10^{-3}~\mathrm{eV}^2 $ & NO  & 2.485 & 2.453 -- 2.514 & 2.419 -- 2.547 & 2.389 -- 2.578 & 1.3 \\
                                      & IO  & 2.465 & 2.434 -- 2.495 & 2.404 -- 2.526 & 2.374 -- 2.556 & 1.2 \\
\hline%---------------------------------------------------------------------
$\sin^2 \theta_{13}/10^{-2}$ & NO & 2.22 & 2.14 -- 2.28 & 2.07 -- 2.34 & 2.01 -- 2.41 & 3.0 \\
                             & IO & 2.23 & 2.17 -- 2.30 & 2.10 -- 2.37 & 2.03 -- 2.43 & 3.0 \\
\hline%---------------------------------------------------------------------
$\sin^2 \theta_{23}/10^{-1}$ & NO & 5.45 & 4.98 -- 5.65 & 4.54 -- 5.81 & 4.36 -- 5.95 & 4.9 \\
                             & IO & 5.51 & 5.17 -- 5.67 & 4.60 -- 5.82 & 4.39 -- 5.96 & 4.7 \\
\hline%---------------------------------------------------------------------
$\delta/\pi$ & NO & 1.28 & 1.10 -- 1.66 & 0.95 -- 1.90  &  0 -- 0.07 $\oplus$ 0.81 -- 2  & 16 \\
             & IO & 1.52 & 1.37 -- 1.65 & 1.23 -- 1.78  &  1.09 -- 1.90  & 9 \\
%\hline%---------------------------------------------------------------------
%$\Delta \chi^2_{\mathrm{{IO}-{NO}}}$ & IO$-$NO & +3.6  \\ [1pt]
%\hline\hline
\end{tabular}
\end{ruledtabular}
%}%end of resizebox
%\vspace*{-.4cm}
\end{minipage}}
\end{table}
%============================================================================

%===========================================================================
\begin{table}[t]
\centering
\resizebox{0.85\textwidth}{!}{\begin{minipage}{\textwidth}
%\captionsetup{width=.9\textwidth}
\caption{\label{Tab:Ordering} %\footnotesize 
Global $3\nu$ analysis of oscillation data. Difference between the absolute $\chi^2$ minima in IO and NO for increasingly rich data sets, including solar, KamLAND (KL), LBL accelerator, SBL reactor, and atmospheric neutrino data. The latter column reports the same difference in terms of $N_\sigma$.}
%\vspace*{0mm}
%\centering
%\resizebox{.9\textwidth}{!}{
\begin{ruledtabular}
\begin{tabular}{lcc}
%\hline\hline
Oscillation dataset & $\Delta \chi^2_{\mathrm{IO}-\mathrm{NO}}$ & $N_\sigma$ \\ 
\hline%---------------------------------------------------------------------
LBL acc.\ + Solar + KL						&  1.8  &  1.3 \\
LBL acc.\ + Solar + KL + SBL reac.			&  5.1 &  2.3 \\
LBL acc.\ + Solar + KL + SBL reac.\ + Atmos. (=~all oscillation data) & 10.0 & 3.2 \\
%\hline\hline
\end{tabular}
\end{ruledtabular}
%}%end of resizebox
%\vspace*{-.4cm}
\end{minipage}}
\end{table}
%============================================================================

Concerning the relative likelihood of IO {\em vs\/} NO, we find that NO is consistently favored in the analysis. Table~\ref{Tab:Ordering} shows that the $\chi^2$ difference between the absolute minima increases by 
enriching the oscillation data set, up to the value $\Delta\chi^2=10.0$ (or $3.2\sigma$) when all data are included. 
Therefore, if the mass ordering information is also marginalized, only the parameter ranges for NO would survive in Table~\ref{Tab:Synopsis}.

\begin{figure}[b]
\begin{minipage}[c]{0.85\textwidth}
\includegraphics[width=1.0\textwidth]{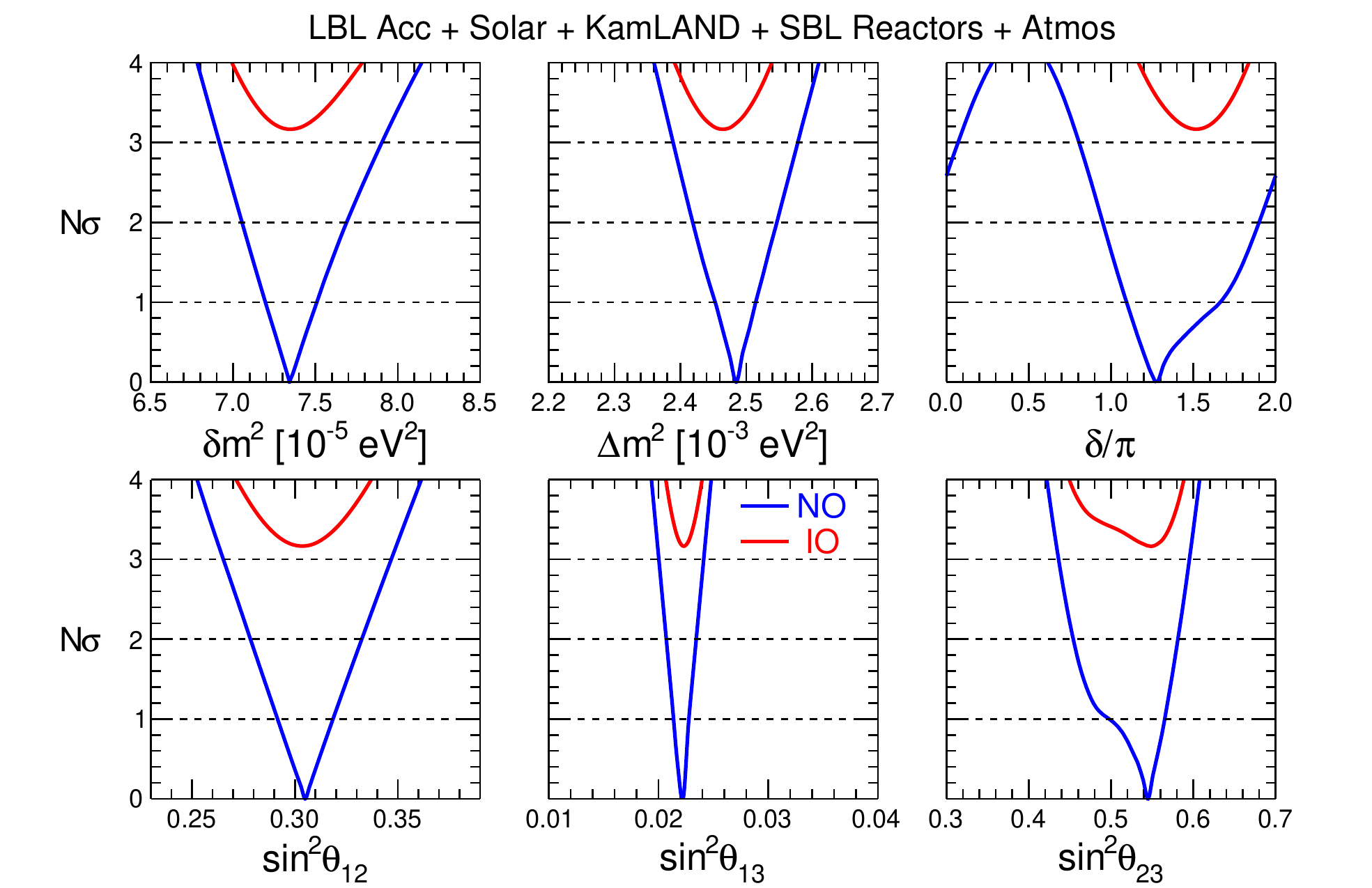}
\caption{\label{Fig:Synopsis}
\footnotesize Global $3\nu$ oscillation analysis. Bounds on the 
parameters $\delta m^2$, $|\Delta m^2|$, $\sin^2\theta_{ij}$, and $\delta$, for NO (blue) and IO (red), in terms of $N_\sigma=\sqrt{\Delta \chi^2}$ from the best fit. In each panel we account for the overall offset 
$\Delta \chi^2_{\mathrm{IO}-\mathrm{NO}}=10.0$, disfavoring the IO case by $3.2\sigma$.   
} \end{minipage}
\end{figure}

Figure~\ref{Fig:Synopsis} reports in graphical form the information about the allowed parameters ranges (Table~\ref{Tab:Synopsis}) and 
about the IO--NO difference (Table~\ref{Tab:Ordering}), including all oscillation data. Our results 
are consistent with those found in recent global analyses \cite{deSalas:2017kay,Esteban:2018azc} and, in particular, are in good agreement 
with the results in \cite{Esteban:2018azc}, except for some differences about the relative likelihood of the two $\theta_{23}$ octants, that
is still ``fragile'' under small changes in the analysis inputs.

%%%%%%%%%%%%%%%%%%%%%%%%%%%%%%%%%%%%%%%%%%%%%%%
\section{Nonoscillation data and constraints}
\label{Sec:Nosc}
%%%%%%%%%%%%%%%%%%%%%%%%%%%%%%%%%%%%%%%%%%%%%%%
  
The previous constraints on the oscillation parameters $(\delta m^2,\,\Delta m^2,\,\theta_{ij})$ reduce the phase space of the three absolute mass observables ($\Sigma,\,m_\beta,\,m_{\beta\beta}$) in both NO and IO \cite{Fogli:2004as}. Moreover, 
as noted, oscillation data disfavor IO at $>3\sigma$. In order to study the sensitivity of nonoscillation data to
the mass ordering, it is useful to proceed by including the oscillation constraints on
$(\delta m^2,\,\Delta m^2,\,\theta_{ij})$ while temporarily ignoring those on the difference $\Delta\chi^2_\mathrm{IO-NO}$,
taken as null instead of $\Delta\chi^2_\mathrm{IO-NO}=10.0$. The latter value will be reintroduced, after completing the
nonoscillation data analysis, in the global data combination.

%%%%%%%%%%%%%%%%%%%%%%%%%%%%%%%%%%%%%%%%%%%%%%%%%%%%%%%%%%%%%%%%%%%%%%%%%%%%%%%%%%%%%%%%%%
%%%%%%%%%%%%%%%%%%%%%%%%%%%%%%%%%%%%%%%%%%%%%%%%%%%%%%%%%%%%%%%%%%%%%%%%%%%%%%%%%%%%%%%%%%

%%%%%%%%%%%%%%%%%%%%%%%%%%%%%%%%%%%%%%%%%%%%%%%%%%%%%%%%%%%%%%%%%%%%%%%%%%%%%%%%%%%%%%%%%%
\begin{figure}[t]
\begin{minipage}[c]{0.8\textwidth}
\includegraphics[width=.78\textwidth]{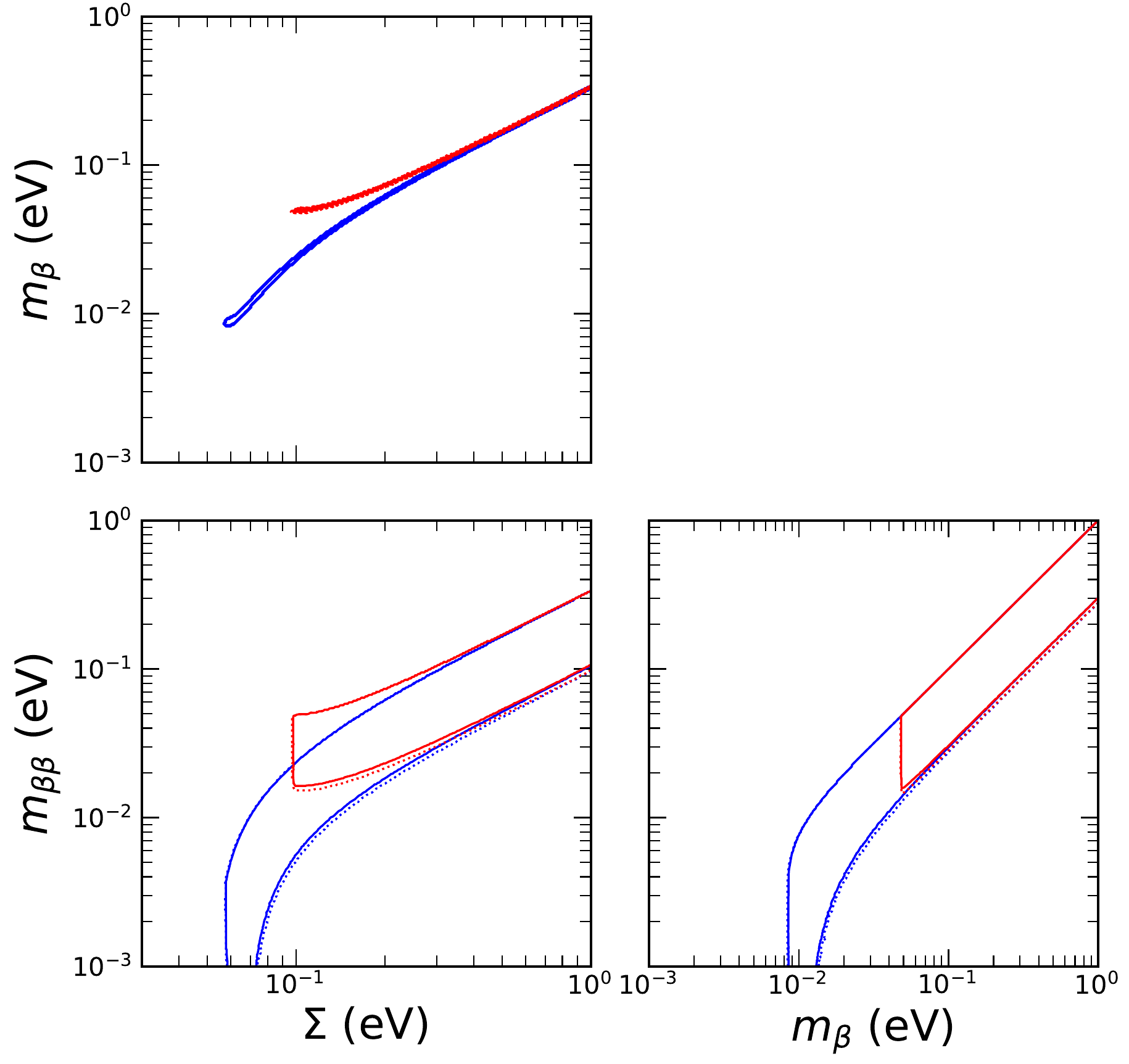}
\caption{\label{Fig:ThreePanels}
\footnotesize 
Oscillation bounds on the nonoscillation observables $(\Sigma,\,m_\beta,\,m_{\beta\beta})$, 
in each of the three planes charted by a pair of such observables. Bounds are shown as contours at
$2\sigma$ (solid) and $3\sigma$ (dotted) for NO (blue) and IO (red) taken separately.  Majorana phases are marginalized away. Note that we take 
$\Delta \chi^2_{\mathrm{IO}-\mathrm{NO}}=0$ in this figure.} 
\end{minipage}
\end{figure}
%%%%%%%%%%%%%%%%%%%%%%%%%%%%%%%%%%%%%%%%%%%%%%%%%%%%%%%%%%%%%%%%%%%%%%%%%%%%%%%%%%%%%%%%%%

Figure~\ref{Fig:ThreePanels} shows the allowed regions for ($\Sigma,\,m_\beta,\,m_{\beta\beta}$) as derived from oscillation data only, in terms of $2\sigma$ and $3\sigma$ bands. The high accuracy achieved in measuring the oscillation parameters
is reflected by the small difference between the 2 and $3\sigma$ contours, as well as by the small width of the bands in the plane charted by the pair ($\Sigma,\,m_\beta)$, not affected by unknown Majorana phases as $m_{\beta\beta}$. 
In this figure we take $\Delta \chi^2_{\mathrm{IO}-\mathrm{NO}}=0$, as discussed above; if the value 10.0 were used, the IO bands  would disappear. 

Let us now discuss the update of nonoscillation data. 
Concerning $m_{\beta\beta}$, a compilation of recent results from neutrinoless double beta decay ($0\nu\beta\beta$) searches has been 
reported and discussed in \cite{Agostini:2019hzm}. In particular, Table~1 therein shows the 90\% C.L.\ upper limits on $m_{\beta\beta}$ 
from single experiments (in terms of their sensitivity for null signal with $m_{\beta\beta}=0$ at best fit), as well as a combined limit $m_{\beta\beta}<66$--$155$~meV at 90\% C.L., 
where the numerical range reflects the spread of nuclear matrix elements in the literature, see 
\cite{Agostini:2019hzm}. For the sake of simplicity
we adopt their median limit, $m_{\beta\beta}<110$~meV at 90\% C.L., corresponding to assume 
$m_{\beta\beta}=0\pm 0.07$~eV ($1\sigma$ error) as $0\nu\beta\beta$-decay input datum in our $\chi^2$ analysis. We note that the
corresponding upper limit at $2\sigma$, $m_{\beta\beta}<0.14$~eV, is slightly stronger than the analogous limit $m_{\beta\beta}<0.18$~eV in our previous analysis \cite{Capozzi:2017ipn}, and reflects the incremental progress in this field.

Concerning $m_\beta$, the KATRIN collaboration recently reported their first and  very promising results, that
can be summarized as: $m^2_{\beta}=-1.0^{+0.9}_{-1.1}$~eV$^2$ at $1\sigma$ \cite{Aker:2019uuj}. By symmetrizing  the lower error 
(unimportant in our parameter space) to match the upper one, we take $m^2_\beta=-1.0\pm0.9$~eV$^2$  as
$\beta$-decay input datum in the $\chi^2$ analysis. A more refined approach using the full likelihood 
profile for $m^2_\beta$ \cite{Aker:2019uuj} is not necessary for the purposes of this Addendum, since the impact of $m_\beta$ 
on neutrino masses is
still weak as compared with that of $0\nu\beta\beta$ or $\Sigma$ (although it will become relevant with future KATRIN data).
In general, as we shall see below, the sensitivity of nonoscillation data to neutrino masses and their ordering is  
dominated by the cosmological constraints on $\Sigma$ and associated variants, so that very 
refined approaches to both $m_\beta$ and  $m_{\beta\beta}$ 
constraints do not really matter (yet).

As in \cite{Capozzi:2017ipn}, we consider a default cosmological model and dataset(s) plus some variants, in order to present constraints ranging from ``aggressive'' to ``conservative'' ones. 
Our default model is the so-called $\Lambda$CDM cosmology augmented with 
$\nu$ masses ($\Lambda$CDM+$\Sigma$), that depends on the following basic parameters : the baryon and the cold dark matter densities $\omega_b$ and $\omega_{cdm}$, the amplitude and tilt of primordial scalar fluctuations $A_s$ and $n_s$, the reionization optical depth $\tau$, and the angular size of the acoustic horizon at decoupling $\theta_\mathrm{MC}$ (see \cite{Cosmo1,Cosmo2,Cosmo3,Cosmo4} for recent reviews). 
Our default dataset includes, in progression, the following experimental inputs:

\begin{itemize}
\item The Planck measurements of Cosmic Microwave Background (CMB) anisotropies from the final 2018 legacy release adopting the same methodology used by the Planck collaboration.  We, therefore, consider a combination of different likelihoods, using the \texttt{commander} likelihood for large scale ($\ell<30$) temperature anisotropies, the \texttt{SimAll} likelihood for large scale polarization anisotropies and the \texttt{Plik} likelihood for temperature, polarization, and cross temperature-polarization anisotropies at small angular scales ($30\le\ell\le2500$). This is the baseline hybrid likelihood used by the Planck collaboration (see \cite{planck2018,plancklike}). In what follows we refer to this dataset as Planck {\scriptsize TT,\,TE,\,EE}. With respect to the Planck 2015 release used in \cite{Capozzi:2017ipn}, the new data is now more reliable in case of the polarization power spectra, with a significant improvement on large angular scales. We, therefore, do not consider anymore the case of Planck temperature alone as in our previous paper \cite{Capozzi:2017ipn}. 
\item The new measurements of the CMB lensing potential power spectrum over multipoles $8\le L \le 400$, also derived from the final Planck 2018 data release \cite{plancklensing}. We refer to this dataset as {``lensing''}. 
\item A compilation of Baryon Acoustic Oscillation (BAO) measurements, given by data from the 6dFGS~\cite{6dFGS}, SDSS MGS~\cite{mgs}, and BOSS DR12~\cite{bossdr12} surveys. We refer to this dataset as {``BAO''}.
\end{itemize}

It should be noted that alternative datasets might provide constraints comparable to our default ones. In particular,
the Ly$\alpha$-forest data from \cite{Yeche:2017upn} would produce, in combination with Planck measurements, a $2\sigma$ bound $\Sigma<0.14$~eV \cite{Yeche:2017upn}. As already noted in  \cite{planck2018}, this bound is close to the one obtained from the Planck+BAO+lensing analysis that, in our case, gives $\Sigma<0.15$~eV (see below). 
In this sense, our default choice of data manages to cover well the typical constraints on $\nu$ masses, as derived 
from current experimental results within the $\Lambda$CDM+$\Sigma$ model. In addition, we have altered the previous default choice, by enlarging either the dataset or the model (with different outcomes on neutrino mass constraints), 
in order to account for some emerging tensions with Planck 2018 data.

 In particular, as additional ``discrepant'' data we 
consider a prior on the Hubble constant as measured by the SH0ES collaboration~\cite{R19} (Riess~{\em et al.} 2019, dubbed {\tt R19}), analysing type-Ia supernovae data from the Hubble Space Telescope using 70 long-period Cepheids in the Large Magellanic Cloud as calibrators. This prior is $H_0=74.03\pm1.42$ km/s/Mpc at $1\sigma$ and we refer to it as $H_0(\mathrm{\tt R19})$. The tension of this prior with Planck 2018 data leads, as we shall see, to tighter constraints on the neutrino mass. We have considered also an alternative  prior on $H_0$ derived from the revised measurement of the Large Magellanic Cloud Tip of the Red Giant Branch extinction from \cite{f20} (Freedman~{\em et al.} 2020, dubbed {\tt F20}), namely,
$H_0(\mathrm{\tt F20})=69.6\pm1.9$ km/s/Mpc, where the quoted statistical and systematics errors have been added in quadrature. We have  
have verified that the combination Planck+R19 covers the range of neutrino
constraints that are obtained by the alternative combinations Planck+F20. Therefore, we shall 
present results for $H_0(\mathrm{\tt R19})$ only, as a paradigmatic example of additional data leading to ``aggressive'' neutrino bounds.

Conversely, some data tensions may be formally relaxed by adding extra degrees of freedom to the model.
In particular, the amount of gravitational lensing in the Planck 2018 CMB spectra is larger than what expected in the $\Lambda$CDM scenario by nearly three standard deviations \cite{planck2018}. 
As in \cite{Capozzi:2017ipn}, we, therefore, extend the $\Lambda$CDM+$\Sigma$ model via an additional parameter $A_\mathrm{lens}$ parameter, that simply rescales the lensing amplitude in the CMB spectra, in order to minimize the effect of this anomaly on the cosmological bounds on the neutrino mass. We refer to this extended scenario as $\Lambda$CDM+$\Sigma$+$A_\mathrm{lens}$. While the constraints obtained in this case on $\Sigma$ are weaker and, therefore, more conservative, it is important to note that the $A_\mathrm{lens}$ parameter is unphysical and that may not properly describe the physical nature of the anomaly. 
 However,
it illustrates a possible ``conservative'' scenario for neutrino mass constraints.

In all cases (default, aggressive, and conservative), 
the cosmological constraints on $\Sigma$ are obtained using the \texttt{CosmoMC} code\cite{cosmomc}, based on a Monte Carlo Markov chain algorithm. Probability posteriors on $\Sigma$ are obtained after marginalization over the remaining nuisance parameters.

%===========================================================================
\begin{table}[t]
%\centering
%\captionsetup{width=.96\textwidth}
\caption{\label{Tab:Cosmo} \footnotesize  
Results of the $3\nu$ analysis of cosmological data.
Our default scenario is based on the standard $\Lambda\mathrm{CDM}+\Sigma$ model
and on Planck 2018
angular CMB temperature power spectrum (TT) plus polarization power spectra (TE, EE), 
with the addition of data from the
 lensing potential power spectrum (lensing) and  Barion Acoustic Oscillations (BAO), separately or in combination
 (cases \#1--3).
A more aggressive scenario is obtained by adding  the
Hubble constant datum from HST observations of Cepheids in the Large Magellanic Cloud  measurements, ${H}_0(\mathrm{\tt R19})$
 (cases \#4--6).
Conversely, a more conservative scenario is obtained by adding an extra degree of freedom ($A_\mathrm{lens}$) to the model
 (cases \#7--9).
For each case we report the $2\sigma$ upper bound on the sum of $\nu$ masses $\Sigma$ (marginalized over NO and IO), together
with the $\Delta\chi^2$ difference between IO and NO, using cosmology only. In the last two columns, we report the 
same information as in the previous two columns, but adding $ m_\beta$ and  $m_{\beta\beta}$ constraints, inducing minor variations. For simplicity, in the text we refer the cases
numbered as 3, 6 and 9 as representative of ``default'', ``aggressive'' and ``conservative''  options, respectively.
}
\vspace*{0mm}
\centering
\begin{ruledtabular}
\begin{tabular}{cllcc|cc}
\multicolumn{3}{l}{Cosmological inputs for nonoscillation data analysis}  & \multicolumn{2}{c|}{Results: Cosmo only} & \multicolumn{2}{c}{Cosmo + $m_\beta$ + $m_{\beta\beta}$}  \\[1mm]
\# & Model & Data set & $\Sigma$ ($2\sigma$)   & $\Delta\chi^2_\mathrm{IO-NO}$ & $\Sigma$ ($2\sigma$)   & $\Delta\chi^2_\mathrm{IO-NO}$ \\[1mm]
\hline%---------------------------------------------------------------------
 0 & $\Lambda\mathrm{CDM}+\Sigma$					& Planck {\scriptsize TT,\,TE,\,EE} 									& $<0.34$ eV & $ 0.9$ & $<0.32$ eV & $ 1.0$   \\
\hline%---------------------------------------------------------------------
 1 & $\Lambda\mathrm{CDM}+\Sigma$					& Planck {\scriptsize TT,\,TE,\,EE} + lensing							& $<0.30$ eV & $ 0.8$ & $<0.28$ eV & $ 0.9$  \\
 2 & $\Lambda\mathrm{CDM}+\Sigma$ 					& Planck {\scriptsize TT,\,TE,\,EE} + BAO								& $<0.17$ eV & $ 1.6$ & $<0.17$ eV & $ 1.7$  \\
 3 & $\Lambda\mathrm{CDM}+\Sigma$ 					& Planck {\scriptsize TT,\,TE,\,EE} + BAO + lensing						& $<0.15$ eV & $ 2.0$ & $<0.15$ eV & $ 2.2$  \\
\hline%---------------------------------------------------------------------
 4 & $\Lambda\mathrm{CDM}+\Sigma$					& Planck {\scriptsize TT,\,TE,\,EE} + lensing + ${H}_0(\mathrm{\tt R19})$	& $<0.13$ eV & $ 3.9$ & $<0.13$ eV & $ 4.0$  \\
 5 & $\Lambda\mathrm{CDM}+\Sigma$ 					& Planck {\scriptsize TT,\,TE,\,EE} + BAO + ${H}_0(\mathrm{\tt R19})$ 		& $<0.13$ eV & $ 3.1$ & $<0.13$ eV & $ 3.2$  \\
 6 & $\Lambda\mathrm{CDM}+\Sigma$ 		& Planck {\scriptsize TT,\,TE,\,EE} + BAO + lensing + ${H}_0(\mathrm{\tt R19})$ \ \	& $<0.12$ eV & $ 3.7$ & $<0.12$ eV & $ 3.8$  \\
\hline%---------------------------------------------------------------------
 7 & $\Lambda\mathrm{CDM}+\Sigma+A_\mathrm{lens}$	& Planck {\scriptsize TT,\,TE,\,EE} + lensing							& $<0.77$ eV & $ 0.1$ & $<0.69$ eV & $ 0.1$  \\
 8 & $\Lambda\mathrm{CDM}+\Sigma+A_\mathrm{lens}$	& Planck {\scriptsize TT,\,TE,\,EE} + BAO								& $<0.31$ eV & $ 0.2$ & $<0.30$ eV & $ 0.3$  \\
 9 & $\Lambda\mathrm{CDM}+\Sigma+A_\mathrm{lens}$	& Planck {\scriptsize TT,\,TE,\,EE} + BAO + lensing						& $<0.31$ eV & $ 0.1$ & $<0.30$ eV & $ 0.2$  \\
\end{tabular}
\end{ruledtabular}
%\vspace*{.6cm}
\end{table}
%============================================================================

%%%%%%%%%%%%%%%%%%%%%%%%%%%%%%%%%%%%%%%%%%%%%%%%%%%%%%%%%%%%%%%%%%%%%%%%%%%%%%%%%%%%%%%%%%
\begin{figure}[b]
\begin{minipage}[c]{0.88\textwidth}
\includegraphics[width=.54\textwidth]{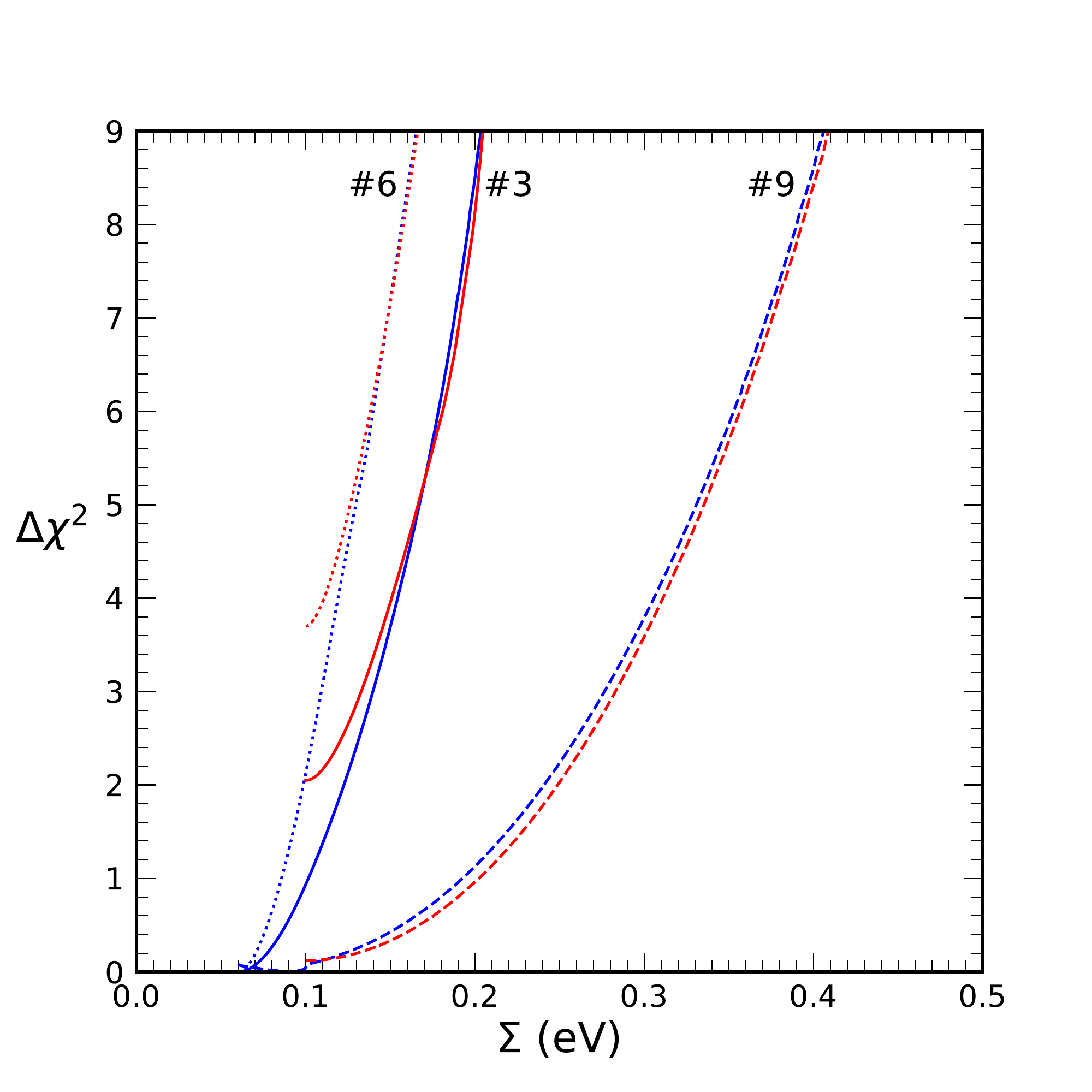}
\caption{\label{Fig:Sigma}
\footnotesize 
$\Delta\chi^2$ curves for NO (blue) and IO (red) from the analysis of cosmological data,  
corresponding to cases numbered in Table~\protect\ref{Tab:Cosmo} as \#6 (left, dotted), \#3 (middle, solid) and \#9 (right, dashed). 
These cases are representative of aggressive, default and conservative options, respectively. Note that, in any case, upper bounds 
on $\Sigma$ can  be placed in the sub-eV range, and that IO is generally disfavored (although only by a tiny amount in the conservative case \#9).} 
\end{minipage}
\end{figure}
%%%%%%%%%%%%%%%%%%%%%%%%%%%%%%%%%%%%%%%%%%%%%%%%%%%%%%%%%%%%%%%%%%%%%%%%%%%%%%%%%%%%%%%%%%
  
In Table~\ref{Tab:Cosmo} we organize the information about cosmological models, input data and fit results as follows.
The first row includes the ``0th'' case with Planck {\scriptsize TT,\,TE,\,EE} data alone. 
The following three rows include our ``default'' options 1--3, where Planck data are combined with either lensing or BAO
inputs or both. In the rows numbered as 4--6, with respect to the cases 1--3 we include the Hubble parameter prior ${H}_0\mathrm{({\tt R19})}$, that
leads to more ``aggressive'' constraints on neutrinos, at the price of introducing 
some tension in the fit. Finally, in the rows numbered as 7--9, 
with respect to the cases 1--3 we allow an extra degree of freedom $A_\mathrm{lens}$ that tends to relax the fit, leading to more
``conservative'' results. In the Table, the fourth and fifth columns show the results of the
cosmological data analysis, in terms of $2\sigma$ upper bounds on $\Sigma$ (marginalized over NO and IO) and $\Delta\chi^2$ difference
between IO and NO. As expected, ``aggressive'' or ``conservative'' options lead to stronger or weaker indications with respect to the ``default'' ones. [We have also replaced the prior $H_0(\mathrm{\tt R19})$ with $H_0(\mathrm{\tt F20})$ (not shown), obtaining less aggressive results, closer to the default ones.] In any case, with respect to our 2017 analysis \cite{Capozzi:2017ipn}, all bounds on $\Sigma$ are now within the sub-eV range, and the overall indication in favor of NO is more
pronounced. These indications remain basically unchanged, or are just slightly corroborated, 
by including subdominant constraints from $\beta$ and $0\nu\beta\beta$ data, as shown in the last two columns. 

In the following two figures, we provide further information complementary to that in Table~\ref{Tab:Cosmo}. For the sake of graphical clarity,
in each group of three cases (1--3, 4--6 and 7--9) we select only the most complete ones (3, 6, and 9) as representative of default,
aggressive and conservative options, respectively.

Figure~\ref{Fig:Sigma} shows the $\Delta\chi^2$ curves for NO and IO (using cosmological data only), with respect to the absolute $\chi^2$ minimum, that
is reached in NO in all cases. One can notice that, in each of the three representative options, the curves tend to converge for increasing values of 
$\Sigma$ as they should, up to
residual differences (not larger than $\delta\chi^2\simeq  0.1$ at any $\Sigma$), 
that quantify the small numerical uncertainty of the analysis. The curves would converge also at small $\Sigma$ in the degenerate approximation
$m_1+m_2+m_3$, that we discard since we do include  the oscillation constraints on $\delta m^2$ and $\Delta m^2$ in the cosmological fit. As
a result, we can correctly quantify the $\chi^2$ differences arising bewteen IO and NO at small values of $\Sigma$, as shown in this figure and numerically reported
in the fifth column of Table~\ref{Tab:Cosmo}.

\begin{figure}[t]
\begin{minipage}[c]{0.90\textwidth}
\includegraphics[width=0.90\textwidth]{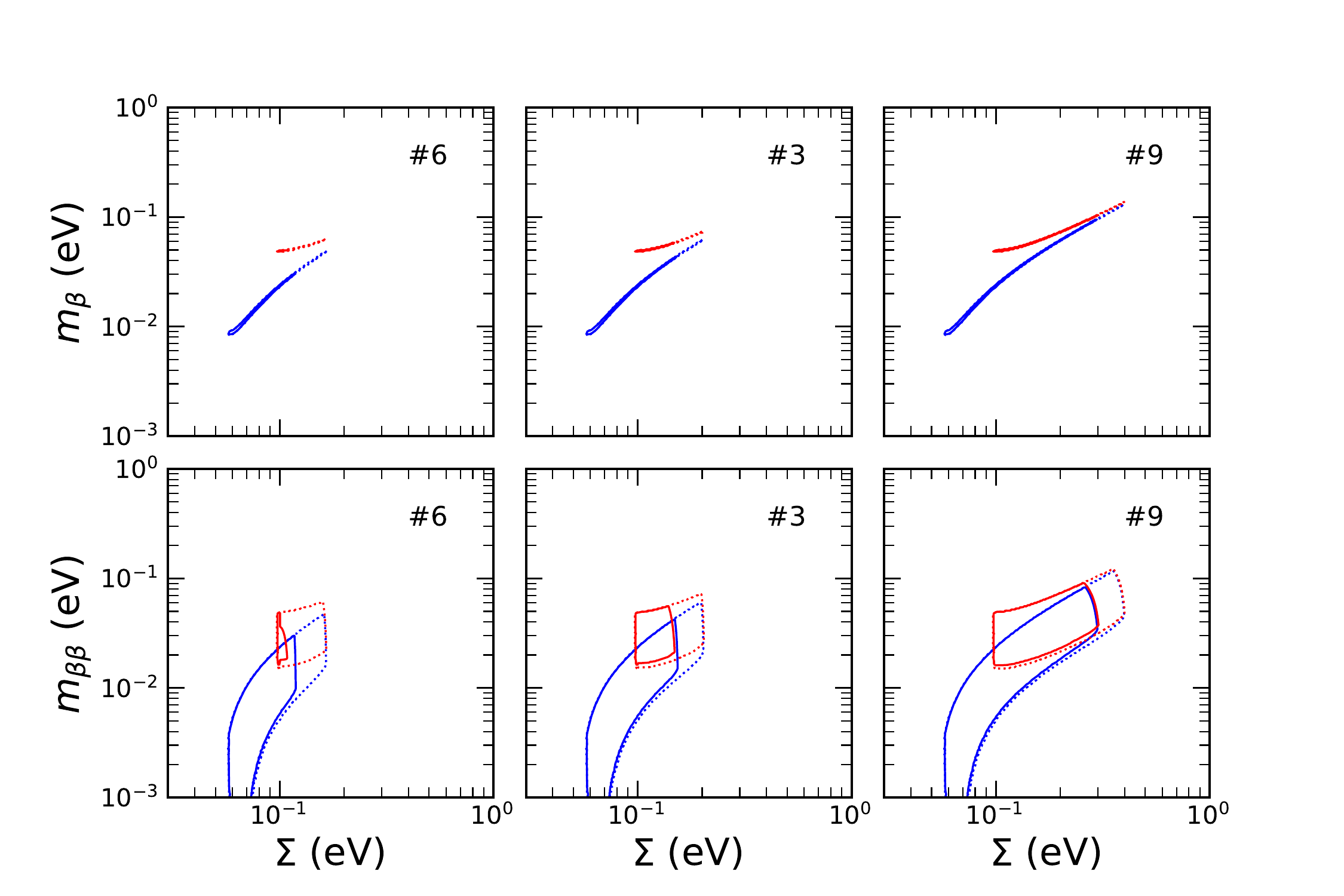}
\caption{\label{Fig:SixPanels}
\footnotesize 
Bounds at $2\sigma$ (solid) and $3\sigma$ (dotted) for NO (blue) and IO (red), as derived by including  nonoscillation data with respect to 
Fig.~\protect\ref{Fig:ThreePanels}, in the upper and lower panels charted by $(\Sigma,\,m_{\beta})$ and by $(\Sigma,\,m_{\beta\beta})$,  
respectively. The bounds include the $\Delta\chi^2$ difference between IO and NO, as reported in the last column of Table~\protect\ref{Tab:Cosmo}.
 The pairs of panels on the left, in the middle and on the right correspond to the cases \#6, \#3 and \#9 in Table~\protect\ref{Tab:Cosmo}, 
 respectively.   } 
\end{minipage}
\end{figure}

Figure~\ref{Fig:SixPanels} shows how the constraints in the planes $(\Sigma,\,m_{\beta\beta})$ and $(\Sigma,\,m_{\beta})$  
are modified (with respect to those in Fig.~\protect\ref{Fig:ThreePanels}) by the fit to nonoscillation data from
cosmology, single and double beta decay. The left, middle and right panels correspond to the cases numbered in Table~\ref{Tab:Cosmo}
as 6 (aggressive), 3 (default) and
9 (conservative), respectively. 
Allowed regions are always present in IO, since nonoscillation data do not yet discriminate IO from NO at $>2\sigma$ in any of the
cases that we have considered. Of course, the IO regions would disappear by adding also the indications in favor of NO derived from oscillation data.

When a direct comparison is possible, our cosmological constraints agree well with the results from similar analyses \cite{hannestad,lattanzi,gariazzo}.  
A Bayesian combination of such constraints with those from $0\nu\beta\beta$ decay has been considered in \cite{Vissani},
where the upper bound $m_{\beta\beta}<0.031$~eV was obtained for $\Sigma <0.14$~eV (at $2\sigma$ for NO). Our closest case in \#3 in Fig.~\ref{Fig:SixPanels}, where we obtain $m_{\beta\beta}<0.04$~eV for $\Sigma <0.15$~eV; the results are in the same ballpark, with secondary differences due to alternative statistical approaches.

%\newpage

%%%%%%%%%%%%%%%%%%%%%%%%%%%%%%%%%%%%%%%%%%%%%%%%%%%%%%%%%%%%%%%%%%%%%%%%%%%%%%%%%%%%%%%%%%
%%%%%%%%%%%%%%%%%%%%%%%%%%%%%%%%%%%%%%%%%%%%%%%%%%%%%%%%%%%%%%%%%%%%%%%%%%%%%%%%%%%%%%%%%%

%%%%%%%%%%%%%%%%%%%%%%%%%%%%%%%%%%%%%%%%%%%%%%%
\section{Synthesis and conclusions}
\label{Sec:Conc}
%%%%%%%%%%%%%%%%%%%%%%%%%%%%%%%%%%%%%%%%%%%%%%%

We conclude this Addendum by merging the information coming from oscillation and nonoscillation data. 
This merging does not alter the bounds on the sum of neutrino masses $\Sigma$ already reported in the sixth column of
Table~\ref{Tab:Cosmo}, and that can be summarized as follows:
\begin{eqnarray}
\Sigma &<& 0.15~\mathrm{eV\ (default)}\ ,\\
\Sigma       &<& 0.12-0.69~\mathrm{eV\ (range)}\ ,
\end{eqnarray}
where we have singled out our default case \#3, and reported the whole range spanned by cases \#0--9, covering variants
more conservative or aggressive than the default one. 
The upper edge of this range corresponds to an effective $\beta$-decay neutrino mass $m_\beta\simeq \Sigma/3\simeq 0.23$~eV, at the sensitivity frontier of
the KATRIN experiment \cite{Aker:2019uuj}.

Concerning the mass ordering discrimination, merging oscillation and nonoscillation data enhance the 
indications in favor of NO, since the $\Delta\chi^2$ contributions in the second 
columns of Table~\ref{Tab:Ordering} and in the last column of Table~\ref{Tab:Cosmo} add coherently. 
The overall indication in favor of NO can be summarized as follows, in standard deviation units:
\begin{eqnarray}
N_\sigma (\mathrm{IO-NO}) &=& 3.5~\mathrm{ (default)}\ ,\\
N_\sigma (\mathrm{IO-NO}) &=& 3.2-3.7~\mathrm{ (range)}\ .
\end{eqnarray}

\begin{figure}[t]
\begin{minipage}[c]{0.90\textwidth}
\includegraphics[width=.90\textwidth]{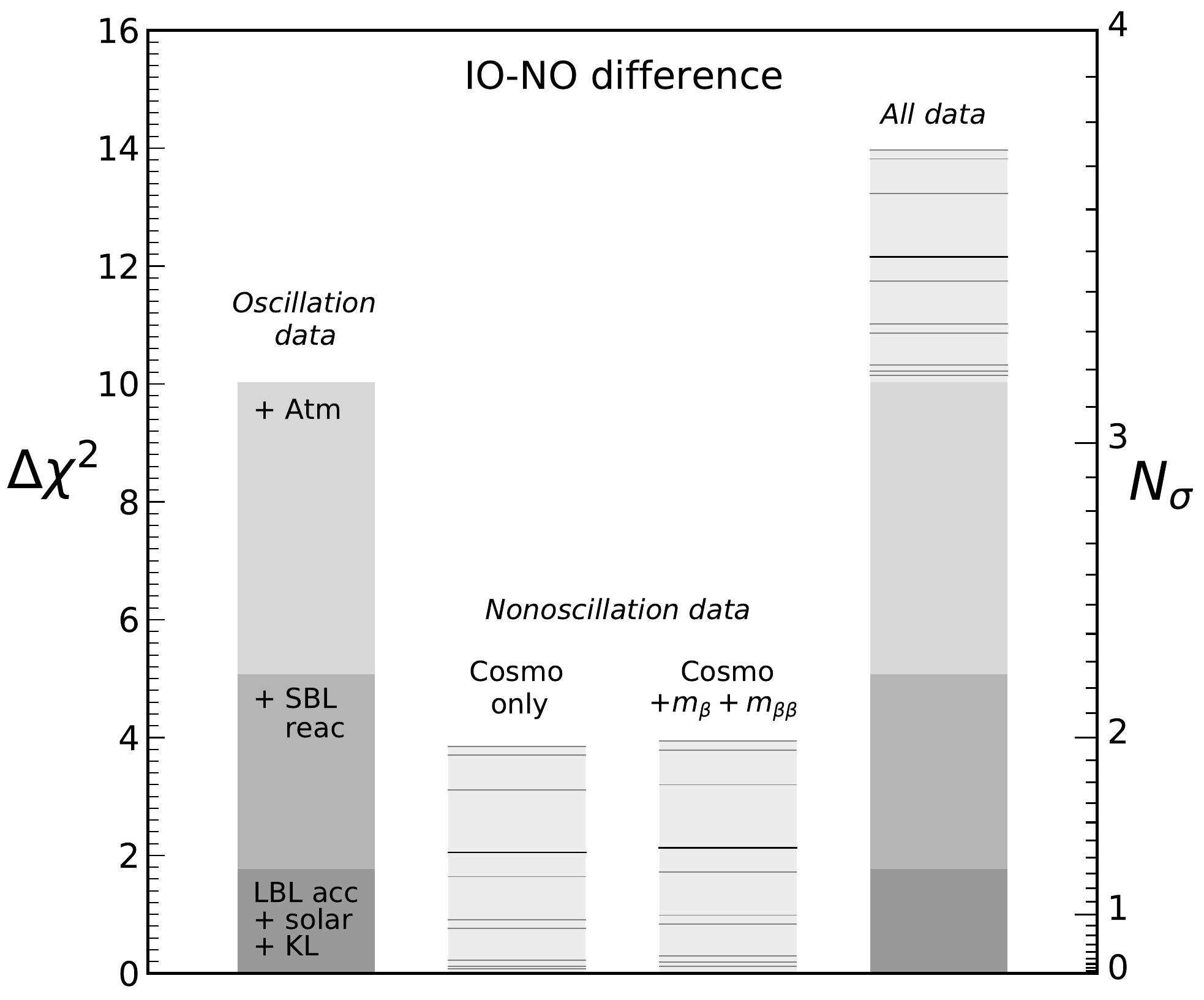}
\caption{\label{Fig:Ordering}
\footnotesize 
Breakdown of contributions to the IO-NO difference from oscillation and nonoscillation data. The latter span a range of cosmological input variants (default, aggressive, and conservative). See the text for details.} 
\end{minipage}
\end{figure}

Figure~\ref{Fig:Ordering} shows the separate and global  contributions to the $\Delta\chi^2(\mathrm{IO-NO})$ difference in graphical form (histogram).
The first bin represents a breakdown of the contributions from oscillation data, as derived in Table~\ref{Tab:Ordering}.  
The second  bin shows the range spanned by
all the cases considered in Table~\ref{Tab:Cosmo}, for the fit to cosmological data only. Each case corresponds to a
horizontal line, with the tick one marking our default case 
\#3.  The third bin shows the slight change induced by adding $m_\beta$ and $m_{\beta\beta}$ constraints, as reported in the last column of Table~\ref{Tab:Cosmo}.
Finally, the fourth bin, obtained by summing the first and third bins, provides the overall indications on mass ordering from
oscillation and nonoscillation data. The vertical axis on the right side translates the results in terms of $N_\sigma$. Although none of the single oscillation or nonoscillation data sets provides compelling evidence for normal ordering yet, their current combination is impressively in favor of this option. 

In conclusion, building upon our previous work \cite{Capozzi:2017ipn}, we have presented improved constraints on absolute neutrino masses and indications on their ordering (favored to be normal), as well as updated bounds on the neutrino oscillation parameters (including hints on the CP phase). In this context, the interplay of oscillation and nonoscillation data remains an important tool to reach a consistent 
picture of neutrino masses and mixings. 

%%%%%%%%%%%%%%%%%%%%%%%%%%%%%%%%%%%%%%%%%%%%%%%%%%%%%%%%%%%%%%%%%%%%%%%%%%%%%%%%%%%%%%%%%%
%%%%%%%%%%%%%%%%%%%%%%%%%%%%%%%%%%%%%%%%%%%%%%%%%%%%%%%%%%%%%%%%%%%%%%%%%%%%%%%%%%%%%%%%%%

%%%%%%%%%%%%%%%%%%%%%%%%%%%%%%%%%%%%%%%%%%%%%%%%%%%%%%%%%%%%%%%%%%%%%%%%%%%%%%%%%%%%%%%%%%%%%%%%%%%%
%%%%%%%%%%%%%%%%%%%%%%%%%%%%%%%%%%%%%%%%%%%%%%%%%%%%%%%%%%%%%%%%%%%%%%%%%%%%%%%%%%%%%%%%%%%%%%%%%%%%

\acknowledgments

This work is partly supported by the Italian Ministero dell'Universit\`a e Ricerca (MUR) through
the research grant no.~2017W4HA7S ``NAT-NET: Neutrino and Astroparticle Theory Network'' under the program PRIN 2017,
and by the Istituto Nazionale di Fisica 
Nucleare (INFN) through the ``Theoretical Astroparticle Physics''  (TAsP) project.
The work of F.C.\ is supported by the Deutsche Forschungsgemeinschaft 
through Grants SFB-1258 ``Neutrinos and Dark Matter in Astro- and Particle Physics (NDM)'' 
and EXC 2094 ``ORIGINS: From the Origin of the Universe to the First Building Blocks of Life''.
The work of E.D.V.\ is supported by the European Research Council through Consolidator Grant no.~681431.
Preliminary results of this work have been presented in various Conferences in 2019-2020.

%%%%%%%%%%%%%%%%%%%%%%%%%%%%%%%%%%%%%%%%%%%%%%%%%%%%%%%%%%%%%%%%%%%%%%%%%%%%%%%%%%%%%%%%%%%%%%%%

\end{document}